\documentclass[twoside]{dis04}
\def\runauthor{Christophe Royon}
\def\shorttitle{Diffractive Higgs production at the LHC}
\begin{document}

\title{DIFFRACTIVE HIGGS PRODUCTION AT THE LHC 
}

\author{Christophe Royon}

\address{DSM-DAPNIA-SPP, CEA Saclay\\
F91 191 Gif-sur-Yvette cedex, France\\
E-mail: royon@hep.saclay.cea.fr }

\maketitle

\abstracts{We use a Monte Carlo implementation of recently developped 
models of double diffractive Higgs production
to assess the sensitivity of the LHC experiments. 
}

\section{Theoretical framework} 
The first proposed  model for $pp \rightarrow p+H+p$, the Bialas-Landshoff (BL)
\cite{bialas} model, 
is based on a summation of two-gluon exchange Feynman graphs coupled to Higgs 
production by the top 
quark loop. 
The non-perturbative character of diffraction at the proton vertices relies 
on the introduction of 
``non-perturbative'' 
gluon propagators which are modeled on the description of soft total 
cross-sections within the additive 
constituent quark model. 

The other popular model for exclusive DPE has been developed by Khoze, Martin, 
Ryskin 
(KMR)~\cite{khoze}. It relies on a purely perturbative, factorized QCD
mechanism applied to 2-gluon exchange among the protons,
without reference to a reggeized Pomeron, and convoluted with the hard
sub-processes $gg\to gg, q\bar{q}$, $H$. The main ingredients of this model 
are the so-called unintegrated 
off-forward 
gluon distributions in the proton.

The survival probability has not been applied in the original computations 
by Bialas et al, and the 
dijet cross-sections 
are found to exceed the CDF experimental bound \cite{cdfdijet}. It has however 
recently been shown, 
using the Good-Walker and Glauber formalisms, 
that the double Pomeron exchange contribution to central diffractive production
of heavy objects has to 
be corrected 
for absorption, in a form determined by the elastic scattering between the 
incident protons
\cite{alex}. 

More details about the theoretical model and its phenomenological
applications can be found in Ref. \cite{ourpap} and \cite{us}. In the following,
we use the BL model for exclusive Higgs production recently implemented in
a Monte-Carlo generator \cite{ourpap}. It has been shown that it gives results
close to the KMR model.

\section{Experimental context}

The analysis is based on a fast simulation of the CMS detector at the LHC
(Similar results would be obtained using the ATLAS simulation).
The calorimetric coverage of the CMS experiment ranges up to a pseudorapidity 
of $|\eta|\sim 5$. 
The region devoted
to precision measurements lies within $|\eta|\leq 3$, with a typical 
resolution on jet energy measurement of $\sim\!50\%/\sqrt{E}$,
where $E$ is in GeV, and a granularity in pseudorapidity and azimuth of 
$\Delta\eta\times\Delta\Phi \sim 0.1\times 0.1$. 

In addition to the central CMS detector, the existence of roman pot detectors
allowing to tag diffractively produced protons,
located on both $p$ sides is assumed \cite{helsinki}. The $\xi$ acceptance and 
resolution have been derived for each device using a complete simulation
of the LHC beam parameters. The combined $\xi$ acceptance is $\sim 100\%$ 
for $\xi$ ranging from $0.002$ to $0.1$, where
$\xi$ is 
the proton fractional momentum loss. The acceptance limit of the device 
closest to the interaction point
is $\xi > \xi_{min}=$0.02. 

In exclusive double pomeron exchange, the mass of the central 
heavy object is given by $M^2 = \xi_1\xi_2 s$ 
\cite{rosto} where $\xi_1$ and $\xi_2$ are
the proton fractional momentum losses measured in the roman pot detectors.

\section{Existence of exclusive events}
\begin{figure}[!thb]
\vspace*{7.0cm}
\begin{center}
\includegraphics{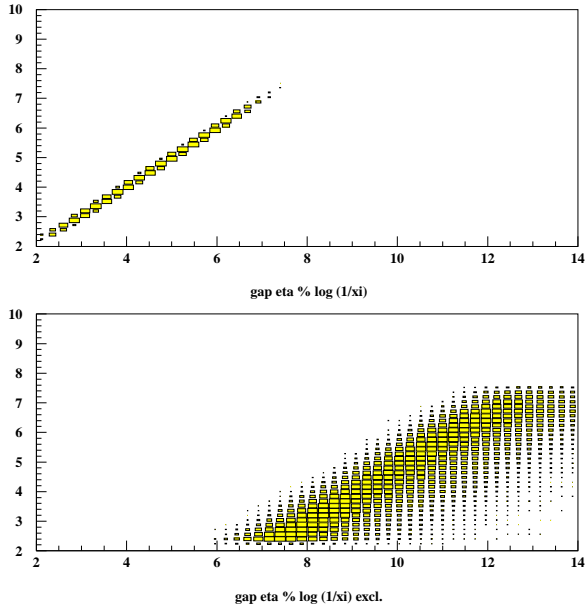}
\vspace{1cm}
\caption[*]{Correlation between the gap size (horizontal axis)
and the value of $log 1/\xi$ measured using tagged protons for inclusive
(upper plot) and exclusive (lower plot) diffractive events. }
\label{gaps}
\end{center}
\end{figure}

The question arises if exclusive events exist or not since they have never been
observed so far. The D\O\ and CDF experiments 
at the Tevatron (and the LHC experiments) are ideal places to look for
exclusive events in dijet or $\chi_C$ channels for instance \cite{cdfdijet}
where exclusive events are expected to occur at high dijet mass
fraction.
So far, no evidence of the existence of exclusive events has been found.
A nice way to show the existence of such events would be to study the
correlation between the gap size measured in both $p$ and $\bar{p}$ directions
and the value of $log 1/\xi$ measured using roman pot detectors, which can be
performed in the D\O\ experiment. The gap size between the
pomeron remnant and the protons detected in roman pot detector 
is of the order of 
$log 1/\xi$ for usual diffractive events (the measurement giving a slightly
smaller value to be in the acceptance of the forward detectors) while
exclusive events show a much higher value for the rapidity gap since the gap
occurs between the jets (or the $\chi_C$) and the proton detected in roman
pot detectors (in other words, there is no pomeron remnant). Fig. \ref{gaps}
shows the correlation between the gap size and $log 1/\xi$ at generator level
for standard diffractive events and exclusive ones
\cite{usbis}
\footnote{To distinguish between pure exclusive and
quasi-exclusive events, other observables such as
the ratio of the cross sections of double diffractive
production of diphoton and dilepton are needed \cite{us}.}. Another observable leading
to the same conclusion would be the correlation between $\xi$ computed
using roman pot detectors and using only the central detector.

\section{Triggering on diffractive Higgs bosons}
Some more details about triggering on diffractively produced Higgs bosons
can be found in Ref. \cite{ourpap}.

At low luminosity ($\sim 10^{33}$ cm$^{-2}$ s$^{-1}$) 
during the first years of the LHC), it is possible to require 
a rapidity gap in the forward region of the calorimeter between
the proton and the jets since the
full available energy is used to produce the Higgs boson in exclusive 
events. This requirement can be performed at the first level of the 
trigger, requiring at the same time the presence of two high $p_T$ jets
in the main detector.

Triggering on diffractively produced Higgs at high luminosity
is not easy since the total
dijet cross-section at the LHC is orders of magnitude too large to allow 
triggering 
on the jets themselves, so benefit must be taken from the specifities of DPE.

If the needed $\xi$ acceptance can be obtained for detectors close enough to 
the interaction point,
requiring two detected protons at the first level trigger eliminates all 
non-diffractive
dijet events and solves the problem. The maximum allowed distance is about 
200m, a number given by the time needed
for a proton to fly from the interaction point to the forward detector, 
for the detector signal to travel back,
and for the trigger decision to be made, within the allowed first level 
trigger latency. This latency is of order 3 $\mu$s 
for the LHC detectors. 
If one requires a proton tag on each side at the first level
of the trigger, this induces a cut on the Higgs mass to be greater
than about 280 GeV.

If one wants to trigger on lower Higgs masses, the trigger is
much more complicated. The first level trigger rate requiring two jets with 
$p_T >$ 40 and 30 GeV, and a 
dijet mass greater than 80 GeV, is 1.1 kHz at low luminosity and 11 kHz at 
high luminosity. 
It is possible to reduce this rate at Level 1 by taking into account the fact 
that 
diffractive jets are more collimated or show less QCD radiation than usual jets
\cite{usbis}.

\section{Sensitivity on standard model Higgs production}
This section summarizes the cuts applied in the analysis.
As said before, both diffracted protons are required to be detected in 
roman pot detectors. The central mass 
is reconstructed using the measurement of $\xi_1$ and $\xi_2$ given
by the forward detectors, giving $M_{miss}= (\xi_1 \xi_2 s)^{1/2}$.

The other cuts are based on detecting well measured, high $p_T$ $b\bar{b}$ 
events.  
We first require the presence of two jets with $p_{T1} >$ 45 GeV, 
$p_{T2} >$ 30 GeV. The difference in 
azimuth between the two jets should be $170 < \Delta \Phi < 190$ degrees, 
asking the jets to be back-to-back. 
Both jets are required to be central, $|\eta| <$  2.5, with the difference
in rapidity of both jets satisfying $| \Delta \eta | < 0.8$. We also apply
a cut on the ratio of the dijet mass to the total mass of all jets
measured in the calorimeters, $M_{JJ}/M_{all} >$ 0.75. 
An additional cut requires a positive $b$ tagging of the jets, eliminating 
all non-b dijet background, with
the efficiency on b-quark dijets quoted above.

The ratio of the dijet mass 
to the missing mass should verify $M_{JJ} / (\xi_1 \xi_2 s)^{1/2}
>$~0.8. This cut requires that all the available Pomeron-Pomeron collision 
energy is used to produce 
the Higgs boson. 

\vspace{-0.5cm}

\section{Results}
Results are given in Fig. 2 for a Higgs mass of 120 GeV, 
in terms of the signal to background 
ratio S/B, as a function of the Higgs boson mass resolution.

In order to obtain an S/B of 3 (resp. 1, 0.5), a mass resolution of about
0.3 GeV (resp. 1.2, 2.3 GeV) is needed. The forward detector design of 
\cite{helsinki} 
claims a resolution of about 2.-2.5 GeV, which leads to a S/B of about 
0.4-0.6. Improvements in this design
would increase the S/B ratio as indicated on the figure.
As usual, this number is enhanced by a large factor if one considers 
supersymmetric Higgs boson 
production with favorable Higgs or squark field mixing parameters.

Our result can be compared to the phenomenological result of
\cite{dkmro}, where experimental issues were addressed within  
the KMR framework. For a missing mass resolution of $\sim$1 GeV, we
have obtained S/B$\sim$1, where the KMR collaboration  
finds S/B$\sim$3. 
In \cite{dkmro}, the background is integrated over a mass window of
1~GeV, assuming that 100\% of the signal lies inside this window. This
is the case only if the mass resolution is significantly smaller   
than 1~GeV, and typically of order 250-300~MeV.
So assuming the result of \cite{dkmro} is given for a gaussian mass
resolution of 1 GeV either underestimates the background by a factor
$\sim$3, or overestimates the signal by the same factor. Taking this
factor into account, and once again assuming that trigger rates and
contamination by inclusive DPE can be kept under control, brings the
KMR estimate to agree with our Monte-Carlo simulation.

\begin{figure}[!thb]
\vspace*{7.0cm}
\begin{center}
\includegraphics{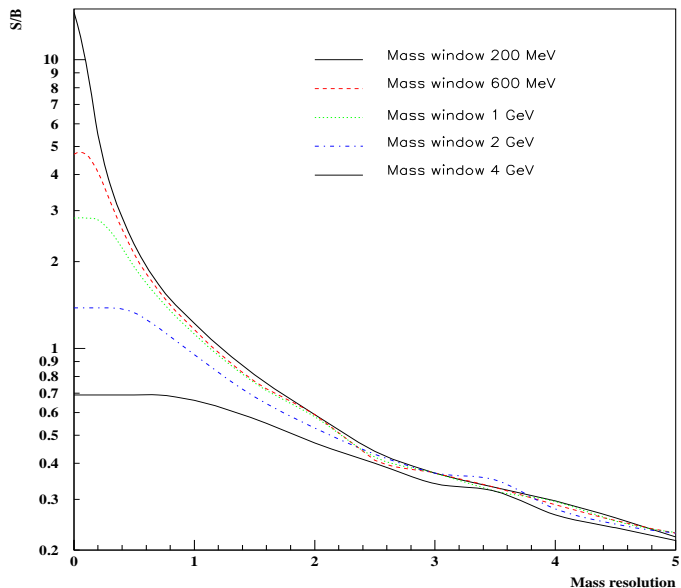}
\vspace{1.5cm}
\caption[*]{Standard Model Higgs boson signal to background ratio as a function of the resolution on 
	the missing mass, in GeV. This figure assumes a Higgs
	boson mass of 120 GeV. }
\end{center}
\end{figure}

\section*{Acknowledgments}
These results come from a fruitful collaboration with 
Maarten Boonekamp and Robi Peschanski.

\end{document}